\documentclass[prl,twocolumn,floatfix,superscriptaddress,amsmath,citeautoscript,aps,longbibliography]{revtex4-2}
\pdfoutput=1
\usepackage[T1]{fontenc}\usepackage[latin1]{inputenc}
\usepackage{dcolumn,bm,csquotes,graphicx,color,booktabs,microtype,afterpage} \graphicspath{{Figures/}}
\usepackage[charter,greekuppercase=italicized]{mathdesign}
\renewcommand{\figurename}{Fig.}
\renewcommand{\tablename}{Table}
\makeatletter\renewcommand{\fnum@figure}[1]{\figurename~\thefigure.}\makeatother
\makeatletter\renewcommand{\fnum@table}[1]{\tablename~\thetable.}\makeatother
\newcount\hh \newcount\mm
\hh=\time \divide\hh by 60
\mm=\hh \multiply\mm by 60 \mm=-\mm
\advance\mm by \time
\def\now{\number\hh:\ifnum\mm<10{}0\fi\number\mm}
\usepackage[colorlinks,plainpages=false,linkcolor=blue,urlcolor=blue,citecolor=blue,pdfpagemode=UseNone,pdfstartview=FitBH]{hyperref}

\newcommand{\zncr}{ZnCr$_\text{2}$Se$_\text{4}$}

\usepackage{soul}
\usepackage{orcidlink} 

\begin{document}
\title{Thermally Enhanced Pseudo-Goldstone Spin Gap in the Isotropic Helimagnet ZnCr$_2$Se$_4$}

\author{N.~D.\ Andriushin\,\orcidlink{0000-0001-5400-8719}}
\affiliation{Institut f{\"u}r Festk{\"o}rper- und Materialphysik, Technische Universit{\"a}t Dresden, D-01069 Dresden, Germany}

\author{A.~Sanders\,\orcidlink{0000-0003-4283-0566}}
\affiliation{Helmholtz-Zentrum Berlin f{\"u}r Materialien und Energie, Hahn-Meitner-Platz 1, 14109 Berlin, Germany}
\affiliation{Dahlem Center for Complex Quantum Systems and Fachbereich Physik, Freie Universit{\"a}t Berlin, Arnimallee 14, 14195 Berlin, Germany}

\author{J.~Reuther\,\orcidlink{0009-0002-3919-7720}}
\affiliation{Helmholtz-Zentrum Berlin f{\"u}r Materialien und Energie, Hahn-Meitner-Platz 1, 14109 Berlin, Germany}
\affiliation{Dahlem Center for Complex Quantum Systems and Fachbereich Physik, Freie Universit{\"a}t Berlin, Arnimallee 14, 14195 Berlin, Germany}

\author{D.~C.\ Peets\,\orcidlink{0000-0002-5456-574X}}
\affiliation{Institut f{\"u}r Festk{\"o}rper- und Materialphysik, Technische Universit{\"a}t Dresden, D-01069 Dresden, Germany}

\author{P.~Steffens\,\orcidlink{0000-0002-7034-4031}}
\affiliation{Institut Laue-Langevin, 71 avenue des Martyrs, CS 20156, 38042 Grenoble Cedex 9, France}

\author{A.~Hiess\,\orcidlink{0000-0002-6457-691X}}
\affiliation{Institut Laue-Langevin, 71 avenue des Martyrs, CS 20156, 38042 Grenoble Cedex 9, France}
\affiliation{European Spallation Source ERIC, P.O. Box 176, 221 00 Lund, Sweden}

\author{J.~Lass\,\orcidlink{0000-0001-6311-9568}}
\affiliation{PSI Center for Neutron and Muon Sciences, Forschungsstrasse 111, 5232 Villigen, PSI, Switzerland}

\author{L.~Prodan\,\orcidlink{0000-0002-2250-8249}}
\affiliation{Experimentalphysik V, Center for Electronic Correlations and Magnetism, Institute of Physics, University of Augsburg, D-86159 Augsburg, Germany}

\author{V.~Tsurkan\,\orcidlink{0000-0001-6001-3151}}
\affiliation{Experimentalphysik V, Center for Electronic Correlations and Magnetism, Institute of Physics, University of Augsburg, D-86159 Augsburg, Germany}
\affiliation{Institute of Applied Physics, Moldova State University, MD 2028 Chisinau, R. Moldova}

\author{V.~Felea}
\affiliation{Institute of Applied Physics, Moldova State University, MD 2028 Chisinau, R. Moldova}

\author{D.~S.\ Inosov\,\orcidlink{0000-0003-0639-2503}}
\thanks{Corresponding author:Dmytro.Inosov@tu-dresden.de}
\affiliation{Institut f{\"u}r Festk{\"o}rper- und Materialphysik, Technische Universit{\"a}t Dresden, D-01069 Dresden, Germany}
\affiliation{W\"urzburg-Dresden Cluster of Excellence on Complexity and Topology in Quantum Matter\,---\,\textit{ct.qmat}, TU Dresden}

\begin{abstract}\noindent\parfillskip=0pt\relax
In magnetically ordered systems, the spin gap in the magnon spectrum is normally expected to decrease on warming and close at the transition out of the long-range order. The presence of a pseudo-Goldstone gap is commonly associated with nonlinear terms in spin-wave theory. However, its precise origin and detailed properties can vary and remain under debate. Here we show by means of inelastic neutron scattering that the pseudo-Goldstone gap in the Heisenberg helimagnet ZnCr$_2$Se$_4$ exhibits an atypical, nonmonotonic temperature dependence: the gap increases over much of the ordered phase before closing near $T_\mathrm{N}$. These results align with recent spin-dynamic simulations based on an effective quartic potential in an isotropic system~[Phys. Rev. B \textbf{113}, L220411 (2026)]. As the pseudo-Goldstone spin gap in ZnCr$_2$Se$_4$ is dominated by quantum effects at subkelvin temperatures, its characteristic growth on warming is caused by thermal fluctuations.
\end{abstract}

\maketitle

\noindent\textit{Introduction.} Ordered states generically have energy gaps\,---\,there should be an energy cost associated with any excitation into an alternative state, otherwise the order would be unstable to even the smallest fluctuations. Near a thermally driven transition, the gap closes as thermal fluctuations increasingly overcome the barrier separating ordered configurations. Nonthermal tuning parameters, such as magnetic field or uniaxial strain, can close this gap through an explicit level crossing between eigenstates. Gaps typically decrease monotonically toward any second-order phase transition into another phase, following characteristic power laws.  The details of how the gap closes as a function of the tuning parameter reveal information on the universality class and key characteristics of the order.

An example of nontrivial gapped excitations is the pseudo-Goldstone mode whose origin is rooted in fluctuations~\cite{PhysRevLett.121.237201, PhysRevResearch.2.043023, PhysRevLett.130.266702, hickey2025universaltemperaturedependentpowerlaw,Rao2025-bx, rdyh-2fqj, PhysRevB.111.184434}. Unlike true Goldstone modes, which stay strictly gapless owing to a continuous symmetry, pseudo-Goldstone modes are gapless only at the harmonic level and acquire a small gap once anharmonic effects are taken into account. They therefore appear as anomalously soft excitations, as reported, for instance, in $\text{Er}_2\text{Ti}_2\text{O}_7$, where an accidental ground-state degeneracy lifted by order-by-disorder has been proposed as the underlying selection mechanism~\cite{PhysRevB.68.020401, PhysRevLett.109.167201, PhysRevLett.109.077204}. However, this is not the only route to a pseudo-Goldstone mode. 

Recently, a general theoretical framework was developed that characterizes the pseudo-Goldstone mode and the conditions under which it appears~\cite{fancelli_anharmonic2026}. Classical calculations show that a system can host such a mode under less strict conditions than those required for the order-by-disorder mechanism. This theoretical framework considers deformations of classical spin states that are associated with an effective quartic potential $V(x)\sim\lambda x^4$\,---\,softer than a harmonic one\,---\,resulting in the stabilization of a pseudo-Goldstone mode with a distinctive temperature scaling. At zero temperature, the classical picture predicts a gapless mode; as the temperature becomes finite and increases, the excitation acquires a gap due to thermal fluctuations. For ordered systems, this situation\,---\,in which the excitation gap grows with temperature\,---\,is highly unusual.

Currently, magnetic materials exhibiting an anharmonic quartic perturbation and consequently hosting the pseudo-Goldstone mode have not yet been established. However, examples are expected among classical isotropic Heisenberg systems with incommensurate spiral ground states related by symmetry~\cite{fancelli_anharmonic2026}. In this Letter, we experimentally examine the pseudo-Goldstone mode in an isotropic system which fulfills all the requirements to exhibit a collective deformation with a quartic potential. As an experimental platform, we choose \zncr, a helimagnet driven by Heisenberg exchange interactions that hosts pseudo-Goldstone modes~\cite{Cameron2016,Inosov_2020,Tymoshenko_2017,zajdel_structure_2017}.

\begin{figure*}
  \includegraphics[width=0.99\linewidth]{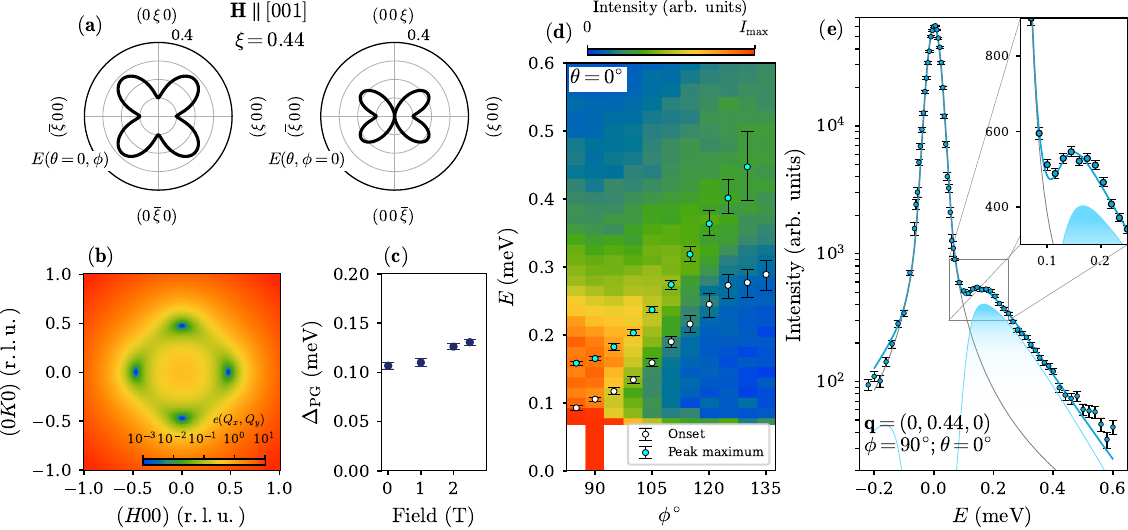}
	\caption{~(a,b)~Dispersion of the low-energy magnon band on a constant  $|\mathbf{q}|$ momentum sphere. $\phi$ is rotation in the [$HK$0] plane orthogonal to the helical propagation vector and $\theta$ is out-of-plane rotation. (b)~The classical energy of a coplanar spin spiral as a function of propagation vector $e(\mathbf{Q}_x, \mathbf{Q}_y)$ in meV units. The energy of interactions is calculated based on values in Ref.~\cite{Tymoshenko_2017}. The ground state energy is taken as zero. (c)~Field dependence of the fitted spin gap (onset) at $T=50$~mK. (d)~The inelastic neutron scattering scans at different rocking angles ($\phi$) across the pseudo-Goldstone mode at zero field and 50~mK after field cooling. Points show the fitted parameters of the inelastic peak. (e)~Background-subtracted inelastic neutron scattering scan, shown on a logarithmic intensity scale, at the pseudo-Goldstone $\mathbf{q}$ vector corresponding to $\phi = 90^\circ$ in panel (d). The inset shows the spin gap on a linear scale. The gray curve is the fitted elastic contribution, the shaded area shows the fitted inelastic peak, and the blue solid curve is the total fit.}
	\label{Fig1}
\end{figure*}

Below 21~K, helical spin order with propagation vector $\mathbf{q} = (0\,0\,\xi)$ is stabilized in \zncr\ through competing exchange interactions and the geometrical frustration of the pyrochlore lattice. Since the three cubic $\langle 1\,0\,0\rangle$ axes are symmetry equivalent, \zncr\ has three coplanar spin-spiral ground states (each with two chiralities) which are symmetry-related. The classical energy of the spiral per spin, $e(\mathbf{Q}_x, \mathbf{Q}_y)$, calculated using exchange parameters from Ref.~\cite{Tymoshenko_2017}, is shown for illustration in Fig.~\ref{Fig1}(b). Once the system settles into one of the ground states, for instance the one with ordering vector $\mathbf{q}_0 = (0\,0\,\xi)$, the spin configuration takes the helical form:
\begin{equation}
\mathbf{S}^0_i = \big(\cos(\mathbf{q}_0\cdot\mathbf{r}_i),\,\sin(\mathbf{q}_0\cdot\mathbf{r}_i),\,0\big).
\end{equation}

The symmetry-related $\mathbf{q}$ vectors that were not selected (in the following called $\mathbf{q_{pG}}$) instead allow for soft spin fluctuations oriented perpendicular to the spin-spiral plane:
\begin{equation}
\mathbf{S}_i = \frac{1}{N_i}\Big[\mathbf{S}^0_i + x\big(0,\,0,\,\cos(\mathbf{q_{PG}}\cdot\mathbf{r}_i)\big)\Big],
\end{equation}
where $x$ is the amplitude of the perturbation and $N_{i}$ is a normalization factor ensuring preservation of spin magnitude. Such spin fluctuations would inevitably exist on a soft quartic potential~\cite{fancelli_anharmonic2026}; hence, the spin gap in \zncr\ is expected to exhibit a temperature evolution atypical of conventional ordered systems, which we confirm experimentally in this Letter by means of inelastic neutron scattering.

\noindent\textit{Inelastic neutron scattering in \zncr.}
The Goldstone mode at the ordering vectors $\mathbf{q}_0$ = (0\,0\,$\pm$0.44) has four additional minima at $\mathbf{q}_{\text{PG}}$ = (0\,$\pm$0.44\,0) and $\mathbf{q}_{\text{PG}}$ = ($\pm$0.44\,0\,0), corresponding to the pseudo-Goldstone modes [see a schematic diagram of the angular dependence of the gap in Fig.~\ref{Fig1}(a,b)].

A small spin gap of order $\sim$0.1 meV is attributed to the pseudo-Goldstone mode. However, it cannot be resolved under normal conditions: all possible magnetic domains have equal population, leading to overlap of the pseudo-Goldstone mode with Goldstone modes from the other domains. This can be overcome by driving the sample into a single-domain state. An external magnetic field favors the helix propagating along the field direction, which can can acquire a small conical component in the field, while raising the energy of the others, allowing the selection of a single magnetic domain (although two chiralities are still possible). Hence the spin dynamics of \zncr\ can be studied in a single-domain state, so long as rethermalization into a multidomain state at higher temperatures is prevented.

In a neutron-scattering experiment on the triple-axis spectrometer ThALES (ILL, France)\,\cite{ThALES_Boehm2008}, the dispersion of the pseudo-Goldstone mode was clearly observed in the single-domain state after cooling in a field of 2~T [Fig.~\ref{Fig1}(d,e)]. A spin gap of $\sim$0.1~meV was resolved at a temperature of 0.05~K in zero field after field cooling. The high energy resolution and background subtraction allowed the spin-gap value to be estimated through fitting (see details of the fit model in Section S1 of the Supplemental Material~\cite{SI}).

Instead of focusing on the absolute value of the pseudo-Goldstone spin gap, we investigate its temperature trend, which can be identified in an experiment more reliably. Measurements capable of resolving the spin gap, similar to those shown in Fig.~\ref{Fig1}(c), must be performed at higher temperatures, where, however, an additional issue appears. The single-domain state prepared by field cooling tends to relax back into a multidomain state under zero-field conditions if temperature fluctuations are strong enough (the characteristic recombination timescale at 8~K is less than an hour~\cite{Inosov_2020}, comparable to the measurement timescale). In order to prevent this relaxation, the temperature dependence may be measured in an applied magnetic field. Importantly, the spin gap does not change significantly below 1~T, as seen in Fig.~\ref{Fig1}(c).

\begin{figure}
  \includegraphics[width=0.99\linewidth]{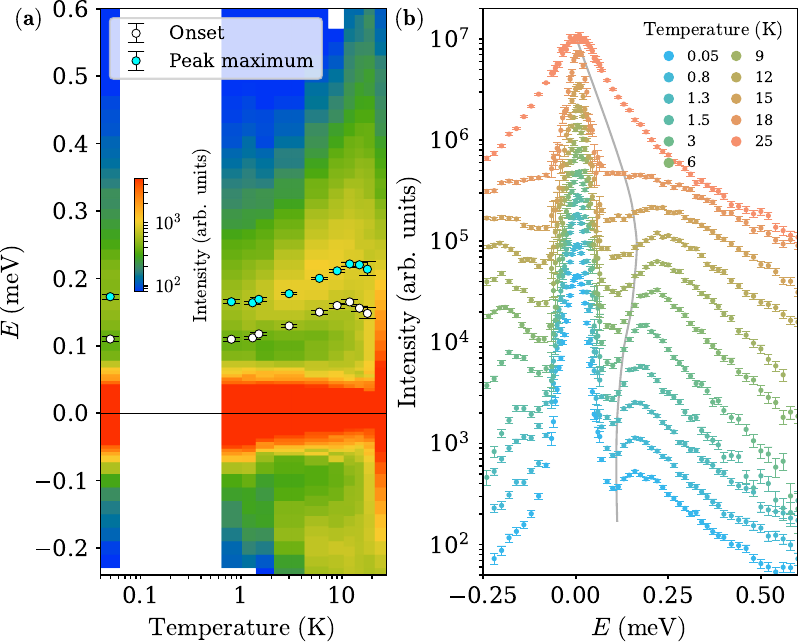}
	\caption{~(a)~Temperature dependence of the background-subtracted inelastic neutron scattering at the pseudo-Goldstone scattering vector in a 1-T field; temperature and intensity are on logarithmic scales, and points show the fit parameters. (b)~The data shown in panel (a) drawn as a waterfall plot on a logarithmic intensity scale. The curves are offset multiplicatively by a factor of 2. The gray line shows interpolation of the fitted onset of the inelastic peak.}
	\label{Fig2}
\end{figure}

The temperature dependence was thus measured in a 1-T field maintaining the selected domain throughout the ordered phase but allowing better data quality at higher temperatures.  The measurement and fit results are summarized in Fig.~\ref{Fig2}. In stark contrast to na\"ive expectations that the energy gap would decrease toward the transition to the paramagnetic state, both the fitted onset and peak maximum of the pseudo-Goldstone mode exhibit {\slshape growth} on warming. The spin gap grows on warming to $\sim$12~K, before closing due to the decrease of the ordered moment and the subsequent transition to the paramagnetic state at $\sim$21~K. At these temperatures, the spin fluctuations are significant and appear in the neutron-scattering signal as a quasielastic contribution [for instance see the $I(E)$ curve at 25~K in Fig.~\ref{Fig2}(b), which is almost fully described by a quasielastic term]. The quasielastic signal persists at temperatures above 30~K~\cite{zajdel_structure_2017} and also exhibits broadening in momentum transfer (details shown in Section~S3 of the Supplemental Material~\cite{SI}). Below $\sim$2~K, the pseudo-Goldstone gap appears to plateau, with no significant further changes.

The bare quartic-potential picture predicts monotonic growth of the pseudo-Goldstone gap, based on temperature fluctuations and an isotropic model. The spectra in the intermediate-temperature regime indeed show a steady increase of the spin gap with temperature. This experimental result confirms the theoretical predictions on the temperature evolution of the pseudo-Goldstone gap and reveals that the spin gap in \zncr\ is associated with oscillations in an anharmonic effective potential. At higher temperatures as the system approaches the paramagnetic state, the trend is gradually replaced by the natural softening observed in conventional magnets. 

\noindent\textit{Oscillations in effective potential.} To theoretically model the temperature dependence of the pseudo-Goldstone gap, one may describe the collective coordinate $x$ associated with the out-of-plane spin deformation at $\mathbf{q}_{\text{PG}}$ as an oscillator in the effective potential~\cite{fancelli_anharmonic2026}
\begin{equation}
V_\text{eff}(x,T)\sim \lambda x^4+\alpha T x^2+\Delta_0^2x^2.\label{eq:veff}
\end{equation}
The first term is the quartic potential associated with a deformation at $\mathbf{q}_{\text{PG}}$. The second term is an entropic, temperature-dependent contribution that arises from integrating out the spin-wave modes to which the quartic mode couples anharmonically. The last term collects all temperature-independent contributions, i.e.\ all effects that leave the mode gapped as $T\rightarrow0$. Candidates are quantum fluctuations of relative order $1/S$ and anisotropic two-body spin interactions. We note that the anisotropic interactions able to contribute to $\Delta_0$ while keeping the helical structure of the spiral state intact are highly constrained and require an explicit breaking of the pyrochlore lattice symmetries~\cite{Andriushin_in_prep}. ZnCr$_2$Se$_4$ is indeed known to undergo a structural distortion to a tetragonal space group below 21~K~\cite{zajdel_structure_2017}. However, the associated deviation from cubic symmetry is very small, with $(\sqrt2a_\mathrm{tet}-c_\mathrm{tet})/(2a_\mathrm{tet}) \sim 6 \times 10^{-4}$. We therefore expect the corresponding contribution to $\Delta_0$ arising from anisotropic interactions to be negligible. We conclude that quantum fluctuations constitute the dominant contribution to $\Delta_0$, consistent with the results of Ref.~\cite{Tymoshenko_2017}, where non-linear corrections to spin-wave theory yielded a good quantitative estimate of $\Delta_0$.

\begin{figure}
  \includegraphics[width=0.99\linewidth]{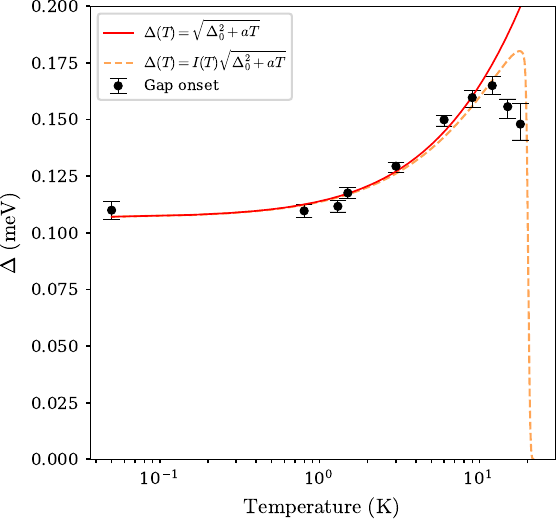}
	\caption{~Fit of the pseudo-Goldstone spin gap temperature evolution. The orange dashed curve is the fit multiplied by the order parameter (see Section S2 of Supplemental Material~\cite{SI}).}
	\label{Fig3}
\end{figure}

As shown in Ref.~\cite{fancelli_anharmonic2026}, the quartic coupling scales as $\lambda\sim1/N$ with the number of spins $N$, so that for a macroscopic sample the regime in which the bare quartic oscillation controls the gap is pushed to inaccessibly low temperatures. The key driver of the temperature dependence is therefore the entropic term. Together with the temperature-independent term $\Delta_0^2 x^2$ this harmonic oscillation gives rise to the gap
\begin{equation}
\Delta(T)=\sqrt{\Delta_0^2+\alpha T}.\label{eq:gap}
\end{equation}
Equation~(\ref{eq:gap}) describes the measured gap well over a wide temperature range, as shown in Fig.~\ref{Fig3}. The subkelvin plateau corresponds to the regime $\alpha T \ll \Delta_0^2$, while the growth above $\sim$2~K reflects the thermally generated gap. The deviations for $T\gtrsim 12 $~K, as the transition is approached, are expected since the description in Eq.~(\ref{eq:veff}) presupposes a well-established spiral at the selected wave vector $\mathbf{q}_0$. The ordered moment of the spiral state, however, vanishes at $T_\mathrm{N}$. A rough estimate of the softening associated with the transition into the paramagnetic state is obtained by multiplying the resulting fit with the order parameter curve, taken as the elastic intensity measured under zero-field cooling. The result is shown as the orange dashed line.

\noindent\textit{Conclusion.}
In this Letter, we presented strong evidence that the spin gap in the pseudo-Goldstone mode in the spinel \zncr\ is enhanced by thermal fluctuations. By means of inelastic neutron scattering, the temperature dependence of the spin gap was revealed, exhibiting an increasing trend on warming, in contrast to conventional monotonic spin-gap behavior following an order parameter. The spin gap approximately follows a $\sqrt{T}$ dependence at intermediate temperatures. To our knowledge, \zncr\ provides the never previously observed experimental realization of this mechanism and can be treated as a starting point for connecting theoretical frameworks focused on pseudo-Goldstone modes with their experimental realization.

The conclusions drawn from the example of \zncr\ are generally applicable to isotropic magnets with symmetry-related planar spin-spiral degenerate ground states. Another plausible platform for hosting similar spin dynamics is the half-Heusler compound GdPtBi, which has cubic symmetry and a commensurate ordering vector $\mathbf{q} = (\frac{1}{2}\,\frac{1}{2}\,\frac{1}{2})$~\cite{PhysRevB.101.014417}. The temperature dependence of the pseudo-Goldstone mode in GdPtBi has yet to be investigated; it has a similar gap value of $\sim$0.15~meV, and its ordering temperature, $T_\mathrm{N} = 9$~K, is lower than that of \zncr. The Gd$^{3+}$ ions have spin $S = 7/2$, which is closer to the classical limit. Hence, if similar behavior is confirmed, the temperature range dominated by thermal effects may be broader in GdPtBi.

\smallskip
\noindent\textit{Acknowledgments}
We acknowledge financial support from the German Research Foundation (DFG) through: individual grants PE~3318/3, IN~209/7, and IN~209/9 (Projects Nos. 455319354, 285734972, and 434257385); project C03 of the Collaborative Research Center SFB~1143 (project No.\ 247310070); through the W\"urzburg-Dresden Cluster of Excellence on Complexity and Topology in Quantum Materials\,---\,\textit{ct.qmat} (EXC~2147, Project No.~390858490); and through the Collaborative Research Center TRR~80 (Project No.~107745057). V.T., L.P., and V.F. acknowledge support by the DFG through TRR~360 (Project No.~492547816) and project ANCD (code 011201, Moldova). This work is based on experiments performed at the Swiss spallation neutron source SINQ, Paul Scherrer Institute, Villigen, Switzerland (proposal No.~2024\,0892)~\cite{Lass_2023}, and at the Institut Laue-Langevin (ILL), Grenoble, France~\cite{4-01-1866, 4-01-1912}.

\noindent\textit{Data availability}
The data that support the findings of this Letter are openly available\,\cite{4-01-1866,4-01-1912}.

\bibliography{bibliography}

\vfill
\onecolumngrid\clearpage
\renewcommand{\thefigure}{S\arabic{figure}}
\renewcommand{\thetable}{S\arabic{table}}
\renewcommand\theequation{S\arabic{equation}}
\renewcommand\thesection{S\arabic{section}}

\centerline{\large{\bf {Supplemental Material to accompany the article}}} 

\vskip1mm

\centerline{\large{\bf {``Thermally Enhanced Pseudo-Goldstone Spin Gap in the Isotropic Helimagnet ZnCr$_2$Se$_4$''}}}

\vskip4mm

\centerline{N.~D.\ Andriushin, A.~Sanders, J.~Reuther, D.~C.\ Peets, P.~Steffens} 

\centerline{A.~Hiess, J.~Lass, L.~Prodan, V.~Tsurkan, V.~Felea, D.~S.\ Inosov} 

\vskip10mm
\twocolumngrid

\section{Low-energy neutron scattering fitting} \label{app:INSfit}

\subsection{General model}

The measured intensity $I(E)$ is described as the sum of an elastic contribution, a gapped inelastic (magnon) contribution, and, where relevant, a quasielastic contribution, convoluted with the instrumental resolution function $F_\mathrm{res}(E)$, plus an incoherent background:
\begin{equation} \label{eq:fit}
\begin{split}
    I(E) ={}& \big(S(E) * F_\mathrm{res}(E)\big) + I_\mathrm{incoh}\text{, where}\\
    S(E) ={}& S^{E}\delta(E) + S_\mathrm{ine}(E) \;[+\, S_\mathrm{qe}(E)]\text{.}
\end{split}
\end{equation}
The elastic term $S^{E}\delta(E)$, once convoluted with $F_\mathrm{res}$, reproduces the instrument resolution profile with amplitude $S^{E}$.

Gapped inelastic contributions are often assumed to be Lorentzian peaks convoluted with instrumental resolution with its intrinsic width representing the excitation damping. However, this approach may be not optimal when dispersion of the excitation is very steep and the profile accumulates a pronounced tail at energies far above the gap due to finite instrument momentum resolution. The energy resolution of the ThALES triple-axis spectrometer reaches tens of microelectronvolts and its momentum resolution is around $\delta|Q| \approx 0.2\,\text{\AA}^{-1}$. However, it is apparent that, in the data collected on \zncr, the spin gap is very small, dispersion is steep and the inelastic scattering data exhibit a long tail which appears approximately linear on a logarithmic intensity scale (and thus cannot be explained by Gaussian or Lorentzian models). Such data still can be phenomenologically described with a model addressing intensity coming from surrounding $q$ points~\cite{Lenander_2026}.

\begin{figure}
\includegraphics[width=0.9\linewidth]{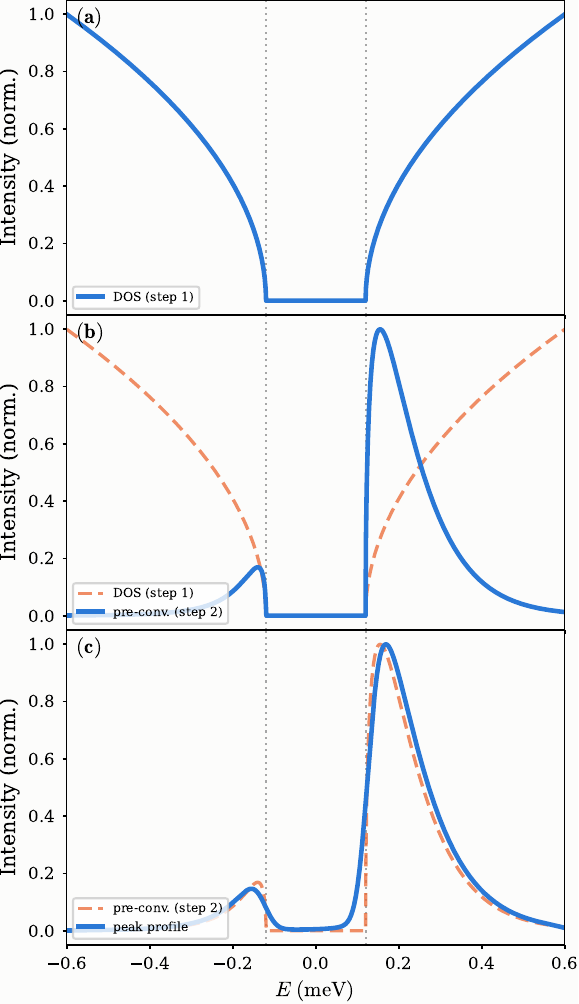}
        \caption{~Calculation of the inelastic peak profile; $\Delta = 0.12, T = 0.8, \sigma_\mathrm{mag} = 0.2, \sigma_r = 0.02, \gamma_r = 0.003$. (a)~Combination of root and Heaviside functions reproducing a density-of-states of quadratic dispersion mode with the edge $\Delta$. (b)~$S_\mathrm{ine}(E)$ function before convolution with instrumental resolution. (c)~The peak profile after convolution. In (b) and (c), the result from the preceding panel is shown for comparison as a dashed line.} \label{fig:magnon_profile}
\end{figure}

The gapped inelastic contribution $S_\mathrm{ine}(E)$ is modeled as a density-of-states edge appropriate for an isotropic quadratic dispersion near the gap minimum, multiplied by a phenomenological envelope accounting for the finite momentum smearing of the dispersion, and weighted by detailed balance:
\begin{equation} \label{eq:magnon}
\begin{split}
    S_\mathrm{ine}(E) ={}& A_\mathrm{mag}\,\Theta(|E|-\Delta)\sqrt{|E|-\Delta}\times{} \\
    &\Big[\big(1+n(|E|,T)\big)\exp\!\left(-\frac{|E-\Delta|}{2\sigma_\mathrm{mag}^{2}}\right) \\
    &\;+\, n(|E|,T)\exp\!\left(-\frac{|E+\Delta|}{2\sigma_\mathrm{mag}^{2}}\right)\Big]\text{,}
\end{split}
\end{equation}
where $\Theta$ is the Heaviside step function, and
\begin{equation}
    n(E,T) = \frac{1}{\exp(E/k_\mathrm{B}T)-1}
\end{equation}
is the Bose occupation number. The two terms in Eq.~\eqref{eq:magnon} correspond to the energy-loss (Stokes, weight $1+n$) and energy-gain (anti-Stokes, weight $n$) processes respectively, each centered on its own side of the gap ($+\Delta$ and $-\Delta$); since $n(E,T)$ is evaluated at $|E|\geq\Delta>0$ throughout the factor does not diverge. The profile of $S_\mathrm{ine}(E)$ has three varying parameters: the spin gap size $\Delta$ defined as an onset of the density-of-states edge, $\sigma_\mathrm{mag}$ describing the tail broadness, and scaling amplitude $A_\mathrm{mag}$. An example of the peak profile calculation is shown in steps in Figure~\ref{fig:magnon_profile}.

\begin{figure}
\includegraphics[width=0.9\linewidth]{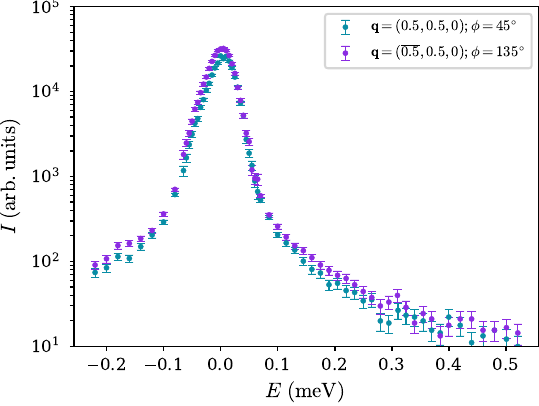}
        \caption{~Measurements of the incoherent contribution. Two data sets represent $\phi = 45^\circ$ and $\phi = 135^\circ$ and have slight differences due to imperfect sample shape and finite misalignment. The background for arbitrary $\phi$ was taken as interpolation of the incoherent contribution.} \label{fig:incoherent}
\end{figure}

The resolution function $F_\mathrm{res}(E)$ is a pseudo-Voigt profile (Gaussian width $\sigma_r$, Lorentzian width $\gamma_r$), normalized to unit height. Resolution parameters $\sigma_r$ and $\gamma_r$ are fit directly to the elastic line of a representative dataset within each experiment. The incoherent background $I_\mathrm{incoh}$ is determined independently from dedicated incoherent-scattering scans, away from any elastic or magnetic excitations (Fig. ~\ref{fig:incoherent}).

A quasielastic component is modeled as a Lorentzian centered at $E = 0$ using the Bose factor multiplied by $E$ to prevent divergence at zero:
\begin{equation} \label{eq:qe}
    S_\mathrm{qe}(E) = A_\mathrm{qe}\,[1 + n(E,T)]\,\frac{E\,\Gamma_\mathrm{qe}}{E^{2}+\Gamma_\mathrm{qe}^{2}}\text{.}
\end{equation}
It was included in the model only for fitting at higher temperatures where the quasielastic contribution is relevant.

\subsection{Low-temperature field and temperature series} \label{sec:lowT}

Data taken at base temperature (0.05~K) for four fields (0, 1, 2, and 2.5~T), together with two additional temperatures (0.8 and 1.3~K) at 1~T, were fit jointly. $\sigma_r$, $\gamma_r$, and $E_0$ were shared across all six datasets. $A_\mathrm{mag}$, $\Delta$ and $\sigma_\mathrm{mag}$ were left independent for each dataset. Results are given in Table~\ref{tab:lowT}.

\begin{table}
\caption{~Low-temperature field/temperature series fit results. $\sigma_r = 0.0179(2)$~meV, $\gamma_r = 0.00340(9)$~meV, $E_0 = 0.0010(2)$~meV (shared).} \label{tab:lowT}
\begin{tabular}{l|l|c|c|c}
\toprule
Dataset & Parameter & $-\sigma$ & Best & $+\sigma$ \\
\midrule\midrule
0~T, 0.05~K & $S^{E}$ & $-831$ & 59661 & 836 \\
& $\Delta$~(meV) & $-0.0032$ & 0.1067 & 0.0032 \\
& $\sigma_\mathrm{mag}$~(meV) & $-0.004$ & 0.222 & 0.004 \\
& $A_\mathrm{mag}$ & $-3.3$ & 62.4 & 3.4 \\
\midrule
1~T, 0.05~K & $S^{E}$ & $-720$ & 45192 & 722 \\
& $\Delta$~(meV) & $-0.004$ & 0.110 & 0.004 \\
& $\sigma_\mathrm{mag}$~(meV) & $-0.005$ & 0.228 & 0.005 \\
& $A_\mathrm{mag}$ & $-4$ & 58 & 4 \\
\midrule
2~T, 0.05~K & $S^{E}$ & $-428$ & 24880 & 431 \\
& $\Delta$~(meV) & $-0.0028$ & 0.1261 & 0.0031 \\
& $\sigma_\mathrm{mag}$~(meV) & $-0.005$ & 0.264 & 0.005 \\
& $A_\mathrm{mag}$ & $-1.7$ & 34.1 & 1.8 \\
\midrule
2.5~T, 0.05~K & $S^{E}$ & $-330$ & 20468 & 330 \\
& $\Delta$~(meV) & $-0.0032$ & 0.1306 & 0.0032 \\
& $\sigma_\mathrm{mag}$~(meV) & $-0.006$ & 0.306 & 0.006 \\
& $A_\mathrm{mag}$ & $-0.9$ & 21.1 & 1.0 \\
\midrule
1~T, 0.8~K & $S^{E}$ & $-681$ & 45914 & 675 \\
& $\Delta$~(meV) & $-0.0031$ & 0.1097 & 0.0029 \\
& $\sigma_\mathrm{mag}$~(meV) & $-0.005$ & 0.228 & 0.005 \\
& $A_\mathrm{mag}$ & $-3.5$ & 57.5 & 3.6 \\
\midrule
1~T, 1.3~K & $S^{E}$ & $-643$ & 45914 & 644 \\
& $\Delta$~(meV) & $-0.0025$ & 0.1117 & 0.0024 \\
& $\sigma_\mathrm{mag}$~(meV) & $-0.004$ & 0.229 & 0.004 \\
& $A_\mathrm{mag}$ & $-3.1$ & 55.8 & 3.2 \\
\bottomrule
\end{tabular}
\\[2pt]
\end{table}

\subsection{High-temperature series} \label{sec:highT}
A separate experiment without a dilution cryostat covered temperatures from 1.5 to 25~K at 1~T. Because the experimental configuration was not identical to the millikelvin-temperature setup, a separate estimation of the background and resolution parameters was performed for data above 1.5~K. $\sigma_r$, $\gamma_r$, and $E_0$ were determined from the lowest-temperature dataset and shared across the series. The incoherent background was likewise determined from a single reference scan specific to this experiment. Results are given in Table~\ref{tab:highT}.

The global fit yields clear trends of a monotonically decreasing elastic peak contribution and a non-monotonic spin-gap dependence with heating. A quasielastic component [Eq.~\eqref{eq:qe}] was included in addition to the magnon term, with both terms present simultaneously. The quasielastic peak is necessary at higher temperatures, but at low temperatures, it becomes poorly constrained because its contribution is almost negligible. Since the onset of the quasielastic peak occurs as a continuous crossover, the specific temperature at which it should be introduced is impossible to pinpoint. Hence, it was decided to include the quasielastic peak across the entire temperature range between 1.5 and 25~K to provide a unified model that preserves parameter trends as well as possible.

At 25~K, \zncr\ becomes paramagnetic, and the spin excitations are entirely replaced by spin fluctuations visible as quasielastic scattering. The inelastic peak was excluded for this particular dataset, as the spectrum is well described by a combination of a small elastic line (representing the incoherent contribution, which gains intensity at elevated temperatures) and the quasielastic line.

\begin{table}
\caption{~High-temperature series fit results. $\sigma_r = 0.0173(3)$~meV, $\gamma_r = 0.0024(4)$~meV, $E_0 = 0.0058(2)$~meV (shared).} \label{tab:highT}
\begin{tabular}{l|l|c|c|c}
\toprule
$T$~(K) & Parameter & $-\sigma$ & Best & $+\sigma$ \\
\midrule\midrule
1.5 & $S^{E}$ & $-861$ & 33403 & 1393 \\
& $\Delta$~(meV) & $-0.0024$ & 0.1176 & 0.0024 \\
& $\sigma_\mathrm{mag}$~(meV) & $-0.004$ & 0.203 & 0.004 \\
& $A_\mathrm{mag}$ & $-4$ & 79 & 4 \\
& $A_\mathrm{qe}$ & $-2.9$ & 5.7 & 1.4 \\
& $\Gamma_\mathrm{qe}$~(meV) & $-0.016$ & 0.016 & 0.024 \\
\midrule
3.0 & $S^{E}$ & $-546$ & 29842 & 540 \\
& $\Delta$~(meV) & $-0.0028$ & 0.1294 & 0.0017 \\
& $\sigma_\mathrm{mag}$~(meV) & $-0.005$ & 0.203 & 0.006 \\
& $A_\mathrm{mag}$ & $-3.8$ & 67.1 & 3.1 \\
& $A_\mathrm{qe}$ & $-0.29$ & 2.37 & 0.22 \\
& $\Gamma_\mathrm{qe}$~(meV) & $-0.027$ & 0.084 & 0.020 \\
\midrule
6.0 & $S^{E}$ & $-534$ & 23616 & 537 \\
& $\Delta$~(meV) & $-0.0027$ & 0.1499 & 0.0019 \\
& $\sigma_\mathrm{mag}$~(meV) & $-0.005$ & 0.217 & 0.008 \\
& $A_\mathrm{mag}$ & $-2.5$ & 42.5 & 1.7 \\
& $A_\mathrm{qe}$ & $-0.24$ & 2.45 & 0.16 \\
& $\Gamma_\mathrm{qe}$~(meV) & $-0.025$ & 0.152 & 0.020 \\
\midrule
9.0 & $S^{E}$ & $-612$ & 19549 & 612 \\
& $\Delta$~(meV) & $-0.0044$ & 0.1597 & 0.0030 \\
& $\sigma_\mathrm{mag}$~(meV) & $-0.008$ & 0.231 & 0.011 \\
& $A_\mathrm{mag}$ & $-2.1$ & 26.1 & 1.7 \\
& $A_\mathrm{qe}$ & $-0.26$ & 2.36 & 0.19 \\
& $\Gamma_\mathrm{qe}$~(meV) & $-0.026$ & 0.181 & 0.024 \\
\midrule
12.0 & $S^{E}$ & $-277$ & 15691 & 277 \\
& $\Delta$~(meV) & $-0.004$ & 0.165 & 0.004 \\
& $\sigma_\mathrm{mag}$~(meV) & $-0.013$ & 0.249 & 0.015 \\
& $A_\mathrm{mag}$ & $-1.3$ & 15.1 & 1.2 \\
& $A_\mathrm{qe}$ & $-0.24$ & 2.62 & 0.21 \\
& $\Gamma_\mathrm{qe}$~(meV) & $-0.028$ & 0.235 & 0.026 \\
\midrule
15.0 & $S^{E}$ & $-415$ & 12892 & 415 \\
& $\Delta$~(meV) & $-0.0052$ & 0.1557 & 0.0034 \\
& $\sigma_\mathrm{mag}$~(meV) & $-0.014$ & 0.286 & 0.019 \\
& $A_\mathrm{mag}$ & $-0.7$ & 7.8 & 0.5 \\
& $A_\mathrm{qe}$ & $-0.22$ & 2.48 & 0.18 \\
& $\Gamma_\mathrm{qe}$~(meV) & $-0.028$ & 0.239 & 0.023 \\
\midrule
18.0 & $S^{E}$ & $-366$ & 12782 & 366 \\
& $\Delta$~(meV) & $-0.007$ & 0.148 & 0.009 \\
& $\sigma_\mathrm{mag}$~(meV) & $-0.033$ & 0.293 & 0.030 \\
& $A_\mathrm{mag}$ & $-0.4$ & 3.1 & 0.5 \\
& $A_\mathrm{qe}$ & $-0.17$ & 3.06 & 0.17 \\
& $\Gamma_\mathrm{qe}$~(meV) & $-0.018$ & 0.234 & 0.016 \\
\midrule
25.0 & $S^{E}$ & $-900$ & 1538 & 887 \\
& $\Delta$~(meV) & \multicolumn{3}{c}{---} \\
& $\sigma_\mathrm{mag}$~(meV) & \multicolumn{3}{c}{---} \\
& $A_\mathrm{mag}$ & \multicolumn{3}{c}{0 (fixed)} \\
& $A_\mathrm{qe}$ & $-0.12$ & 5.86 & 0.12 \\
& $\Gamma_\mathrm{qe}$~(meV) & $-0.0021$ & 0.0619 & 0.0021 \\
\bottomrule
\end{tabular}
\end{table}

\subsection{Angular ($\phi$) dependence} \label{sec:phidep}
\begin{figure}
\includegraphics[width=0.9\linewidth]{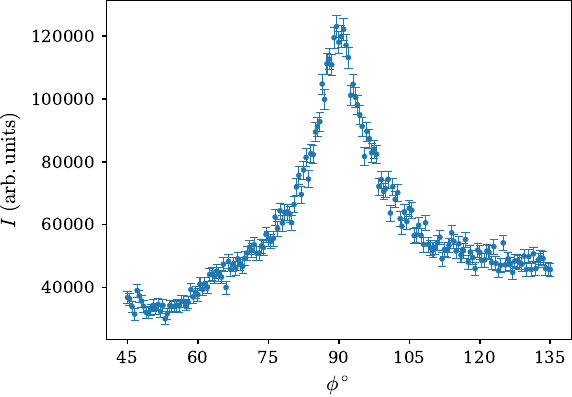}
        \caption{~Rocking scan at $E = 0$ in zero field at the base temperature 0.05~K. $\phi = 90^\circ$ corresponds to the Pseudo-Goldstone reciprocal vector $\mathbf{q} = (0, 0.44, 0)$.} \label{fig:phielastic}
\end{figure}

An additional series of eleven scans at 0~T and 0.05~K, spanning $\phi$~$=85$--$135^\circ$ (where $90^\circ$ corresponds to the pseudo-Goldstone reciprocal vector), was used to map the gap as a function of rocking angle around $\mathbf{Q}=0$. Only one of these scans covered the full energy range including the elastic line; $\sigma_r$, $\gamma_r$, and $E_0$ were determined from that dataset and shared across the series. For the remaining ten datasets, which do not extend below $E\approx0.07$~meV, the elastic amplitude $S^{E}$ cannot be independently fit; it was instead fixed at each angle from a separate elastic-line intensity scan versus $\phi$, scaled by a single proportionality constant determined at the angle common to both measurements (Figure~\ref{fig:phielastic}). $\Delta$, $\sigma_\mathrm{mag}$, and $A_\mathrm{mag}$ were left independent at each angle. Results of individual fits are given in Table~\ref{tab:phidep}.

\begin{figure}
\includegraphics[width=0.9\linewidth]{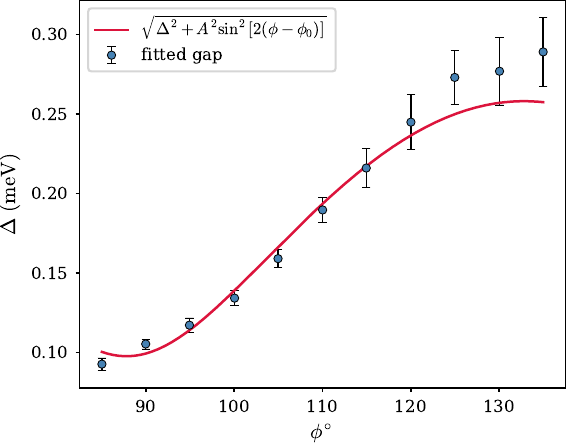}
        \caption{~Angular dependence of the fitted spin gap size.} \label{fig:phidispersion}
\end{figure}

The resulting $\Delta(\mathrm{\phi})$ was compared with the form
\begin{equation} \label{eq:dispersion}
    \Delta(\phi) = \sqrt{\Delta_0^{2} + A^{2}\sin^{2}\!\big(2(\phi-\phi_0)\big)}\text{,}
\end{equation}
giving parameters $\Delta_0 = 0.097$~meV, $A = 0.238$~meV, and $\mathrm{\phi}_0 = 87.8^\circ$ (see Fig.~\ref{fig:phidispersion}).

\begin{table}
\caption{Angular ($\phi$) series fit results. $\sigma_r = 0.0178(3)$~meV, $\gamma_r = 0.0032(2)$~meV, $E_0 = 0.0008(3)$~meV (shared). $S^{E}$ is fixed from the independent elastic-scan calibration (Sec.~\ref{sec:phidep}), not fit.} \label{tab:phidep}
\begin{tabular}{l|l|c|c|c}
\toprule
$\phi$~($^\circ$) & Parameter & $-\sigma$ & Best & $+\sigma$ \\
\midrule\midrule
85.02 & $S^{E}$ (fixed) & \multicolumn{3}{c}{52317} \\
& $\Delta$~(meV) & $-0.0043$ & 0.0927 & 0.0034 \\
& $\sigma_\mathrm{mag}$~(meV) & $-0.0035$ & 0.2364 & 0.0040 \\
& $A_\mathrm{mag}$ & $-2.7$ & 61.5 & 2.7 \\
\midrule
89.98 & $S^{E}$ (fixed) & \multicolumn{3}{c}{60194} \\
& $\Delta$~(meV) & $-0.0034$ & 0.1054 & 0.0028 \\
& $\sigma_\mathrm{mag}$~(meV) & $-0.004$ & 0.222 & 0.004 \\
& $A_\mathrm{mag}$ & $-3.4$ & 63.2 & 3.2 \\
\midrule
94.93 & $S^{E}$ (fixed) & \multicolumn{3}{c}{53805} \\
& $\Delta$~(meV) & $-0.004$ & 0.117 & 0.005 \\
& $\sigma_\mathrm{mag}$~(meV) & $-0.006$ & 0.235 & 0.006 \\
& $A_\mathrm{mag}$ & $-4$ & 50.3 & 4 \\
\midrule
100.04 & $S^{E}$ (fixed) & \multicolumn{3}{c}{41352} \\
& $\Delta$~(meV) & $-0.005$ & 0.134 & 0.005 \\
& $\sigma_\mathrm{mag}$~(meV) & $-0.006$ & 0.244 & 0.007 \\
& $A_\mathrm{mag}$ & $-3.1$ & 43.6 & 3.2 \\
\midrule
104.96 & $S^{E}$ (fixed) & \multicolumn{3}{c}{33407} \\
& $\Delta$~(meV) & $-0.006$ & 0.159 & 0.006 \\
& $\sigma_\mathrm{mag}$~(meV) & $-0.008$ & 0.263 & 0.009 \\
& $A_\mathrm{mag}$ & $-2.8$ & 34.0 & 2.9 \\
\midrule
110.01 & $S^{E}$ (fixed) & \multicolumn{3}{c}{30417} \\
& $\Delta$~(meV) & $-0.008$ & 0.189 & 0.008 \\
& $\sigma_\mathrm{mag}$~(meV) & $-0.013$ & 0.276 & 0.015 \\
& $A_\mathrm{mag}$ & $-3.2$ & 27.5 & 3.3 \\
\midrule
114.95 & $S^{E}$ (fixed) & \multicolumn{3}{c}{29260} \\
& $\Delta$~(meV) & $-0.012$ & 0.216 & 0.012 \\
& $\sigma_\mathrm{mag}$~(meV) & $-0.023$ & 0.309 & 0.027 \\
& $A_\mathrm{mag}$ & $-3.1$ & 20.1 & 3.5 \\
\midrule
120.01 & $S^{E}$ (fixed) & \multicolumn{3}{c}{28185} \\
& $\Delta$~(meV) & $-0.019$ & 0.244 & 0.016 \\
& $\sigma_\mathrm{mag}$~(meV) & $-0.04$ & 0.33 & 0.05 \\
& $A_\mathrm{mag}$ & $-4$ & 17 & 4 \\
\midrule
124.97 & $S^{E}$ (fixed) & \multicolumn{3}{c}{27299} \\
& $\Delta$~(meV) & $-0.019$ & 0.272 & 0.016 \\
& $\sigma_\mathrm{mag}$~(meV) & $-0.04$ & 0.35 & 0.07 \\
& $A_\mathrm{mag}$ & $-3.4$ & 15.1 & 3.9 \\
\midrule
130.04 & $S^{E}$ (fixed) & \multicolumn{3}{c}{26703} \\
& $\Delta$~(meV) & $-0.022$ & 0.276 & 0.020 \\
& $\sigma_\mathrm{mag}$~(meV) & $-0.07$ & 0.41 & 0.14 \\
& $A_\mathrm{mag}$ & $-3.2$ & 12. & 3.8 \\
\midrule
134.98 & $S^{E}$ (fixed) & \multicolumn{3}{c}{26557} \\
& $\Delta$~(meV) & $-0.023$ & 0.289 & 0.020 \\
& $\sigma_\mathrm{mag}$~(meV) & $-0.08$ & 0.42 & 0.17 \\
& $A_\mathrm{mag}$ & $-3.2$ & 11.5 & 3.9 \\
\bottomrule
\end{tabular}
\end{table}

\section{Order parameter fit} \label{app:OrderParameter}

\begin{figure}
\includegraphics[width=0.9\linewidth]{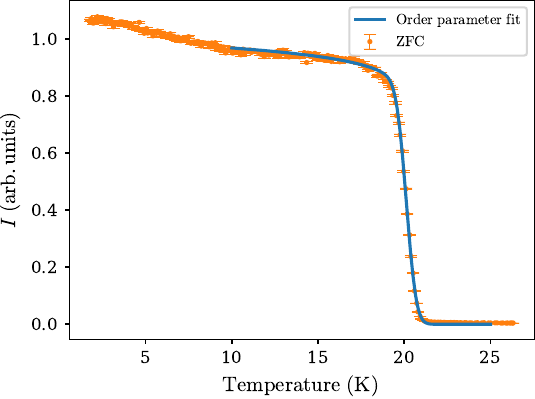}
        \caption{~The elastic intensity measured at ordering vector on zero field cooling. The data is normalized.} \label{fig:orderparameter}
\end{figure}

The elastic neutron scattering signal measured at $\mathbf{q} = (0.44, 0, 0)$ after zero-field cooling shows the temperature dependence of the helical magnetic Bragg intensity in \zncr\ (see Fig.~\ref{fig:orderparameter}). The fitted curve was used as a reference for showing the expected spin gap softening associated with he transition into the paramagnetic state (Fig.~3 of the main text). The transition between the paramagnetic and ordered phases is rather sharp and the magnetic Bragg intensity has an almost step-like temperature dependence near the critical temperature. The measured data were fitted with a power law function with a sharp cutoff, which was smeared by a Gaussian in order to better capture curve features near the transition:

\begin{equation}
    I_0(T; T_0) = \left(1 - \frac{T}{T_0}\right)^{\beta} \Theta(T_0 - T),
    \label{eq:bare}
\end{equation}
convolved with a Gaussian distribution of transition points,
\begin{equation}
    G(T; \sigma) = \frac{1}{\sigma\sqrt{2\pi}}
    \exp\!\left(-\frac{T^2}{2\sigma^2}\right),
    \label{eq:kernel}
\end{equation}
to capture the rounding of the transition seen in the data:
\begin{equation}
    I(T) = \left(I_0 * G\right).
    \label{eq:order_param_smeared}
\end{equation}
where $\Theta$ is the Heaviside step function setting the unsmeared intensity to zero for $T>T_0$; convolution with $G$ accounts phenomenologically for the observed rounding of the transition (i.e.\ the order parameter vanishes once $T$
exceeds a given $T_0'$). The obtained values are $T_0 = 20.150\pm0.002$, $\beta = 0.0459\pm0.0007$ and $\sigma = 0.4462\pm0.0005$.

\section{Multi-domain state dispersions} \label{app:Camea}

\begin{figure}
\includegraphics[width=0.99\linewidth]{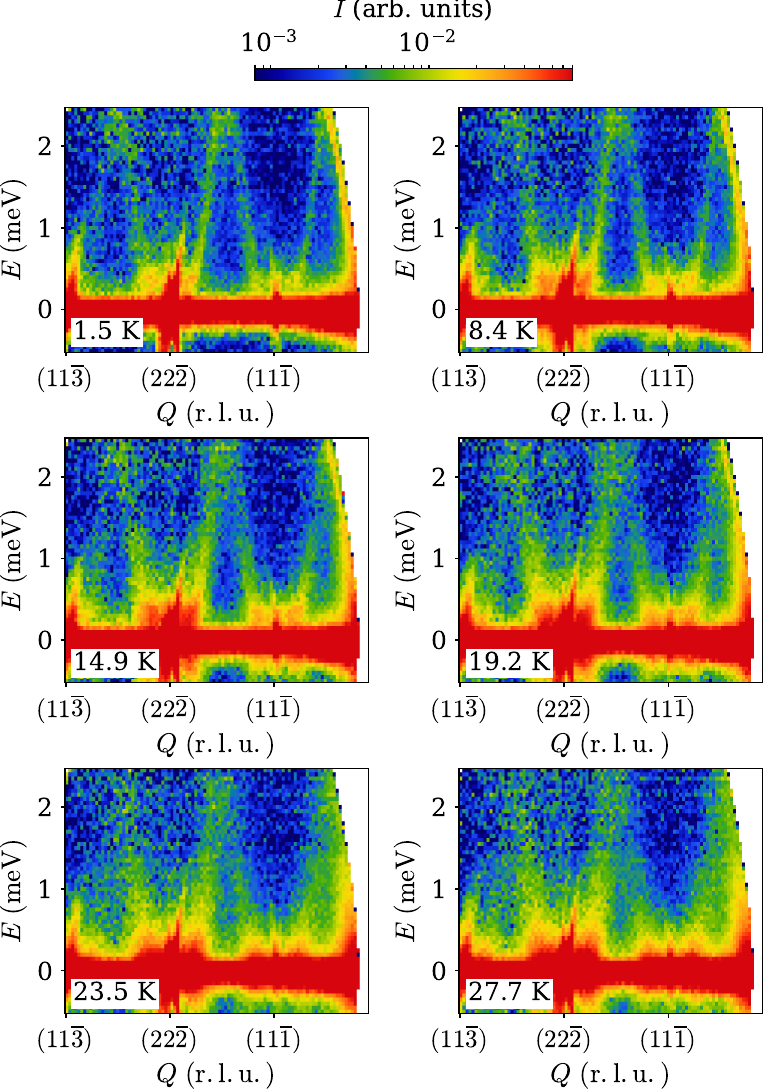}
        \caption{~Dispersion along the [110] reciprocal-space direction in the multidomain state at different temperatures.} \label{fig:camea}
\end{figure}

The magnon dispersions along the [110] reciprocal-space direction were measured in \zncr. The measurement was performed using the multiplexing neutron spectrometer CAMEA, at PSI, Switzerland, with energy transfers covering the relevant range of the spin-wave spectrum. Data were acquired in sample rotation scans covering 80 degrees in steps of 1 degree, measuring ~30 s/point. These scans were measured for tank positions of $-$40, $-$44, $-$74, and $-$78 degrees for incoming energies of 4.5 and 4.629~meV. These were all performed at base temperature, 8.3, 14.9, 19.2, 21.3, 23.5, and 27.7~K in vanishing magnetic field. Figure~\ref{fig:camea} shows the resulting intensity map as a function of momentum and energy transfer along [110]. Since in this experiment the multidomain state of \zncr\ was measured, the inelastic scattering signal combines the pseudo-Goldstone modes coming from two orthogonal domains and the actual Goldstone mode. With increasing temperature, the band broadens progressively in momentum, and above the ordering temperature of 21~K, paramagnon-like excitations persist even though long-range magnetic order has vanished. The dispersions align with data reported in Ref.~\cite{Inosov_2020}. The persistent paramagnon signal was also evident from neutron scattering data measured on a powder sample~\cite{zajdel_structure_2017}.

\end{document}